\documentclass[journal]{IEEEtran}

\usepackage{float}

\usepackage{cite}
\ifCLASSINFOpdf
  \usepackage[pdftex]{graphicx}
\else
  \usepackage[dvips]{graphicx}
\fi
\usepackage{amsmath}
\begin{document}
\title{Visual Entropy and the Visualization of Uncertainty}

\author{Nicolas S. Holliman,\IEEEmembership{ Member,~IEEE Computer Society,}
        Arzu Çöltekin, Sara J. Fernstad, Lucy McLaughlin, Michael D. Simpson, and Andrew J. Woods}% <-this % stops a space

%\thanks{M. Shell was with the Department
%of Electrical and Computer Engineering, Georgia Institute of Technology, Atlanta,
%\thanks{Manuscript drafted July 2021.}% <-this % stops a space


\markboth{King's College London, April 2022}%
{Shell \MakeLowercase{\textit{et al.}}: Bare Demo of IEEEtran.cls for IEEE Journals}




\maketitle

\begin{abstract}
Background—Even though data visualizations (and underlying data) almost always contain uncertainty, it remains complex to communicate and interpret uncertainty representations. Consequently, uncertainty visualizations for non-expert audiences are rare.
%It is possible to find many different visual representations of data values in visualizations, it is less common to see visual representations that include uncertainty, especially in visualizations intended for non-technical audiences. 
Objective—our aim is to rigorously define and evaluate the novel use of visual entropy as a measure of shape that allows us to construct an ordered scale of glyphs for use in representing both uncertainty and value in 2D and 3D environments. Method—We use sample entropy as a numerical measure of visual entropy to construct a set of glyphs using R and Blender which vary in their complexity. Results—An exact binomial analysis of a pairwise comparison of the glyphs shows a majority of participants ($n=87$) ordered each glyph as predicted by the visual entropy score with large effect size (Cohen's $g >0.25$). We also evaluate whether the glyphs effectively represent uncertainty using a signal detection method in a search task. Participants ($n=15$) were able to find glyphs representing uncertainty with high sensitivity and low error rates. Conclusion—visual entropy is a successful novel approach to representing ordered data and provides a channel that allows the uncertainty of a measure to be presented alongside its mean value.
\end{abstract}


% \fi
\IEEEpeerreviewmaketitle

% This stuff fixes a problem with \thanks not working.
%
\newcommand\blfootnote[1]{%
  \begingroup
  \renewcommand\thefootnote{}\footnote{#1}%
  \addtocounter{footnote}{-1}%
  \endgroup
}

\makeatletter
\def\footnoterule{ 
  {\kern-3\p@
  \hspace*{2cm}\vbox{ \hrule \@width 2in \kern 2.6\p@} } }% the \hrule is .4pt high
\makeatother

\blfootnote{N.S. Holliman is with the Department of Informatics, King's College, London, WC2B 4BG. E-mail: nicolas.holliman@kcl.ac.uk}%
\blfootnote{A. Çöltekin is with the Institute for Interactive Technologies, School of Engineering, University of Applied Sciences \& Arts Northwestern Switzerland.}%
\blfootnote{S.J. Fernstad. L. McLaughlin, and M. Simpson are with the Faculty of Science at Newcastle University.}
\blfootnote{A.J. Woods is with the Centre for Marine Science and Technology at Curtin University, Australia.} 

\section{Introduction}

\IEEEPARstart{U}{ncertainty} is a concept that can be complex to present in meaningful, comparable ways to both expert and non-expert audiences. In addition, while it is often seen as valuable to present uncertainty, it is much less often directly depicted in visualizations \cite{hullman2019authors}, possibly due to its complex nature.

Inherently, uncertainty visualization involves a need to comprehend \textit{a lack of a property}, and it is well known that human decision making is both ignorance averse and influenced differentially by the use of negative vs. positive framing of decisions \cite{hardman2009judgment}. 
Outcomes framed using negative concepts such as uncertainty are less often chosen in decision making than positive concepts such as certainty.
In this paper, we introduce a definition of visual entropy that we believe may find uses in a range of visualization applications. We hypothesise visual entropy should relate to visual complexity, so that low visual entropy describes smooth visual signals and high visual entropy describes complex, more disordered, visual signals. This allows us to begin to define a scale of visual entropy, and to implement this, we propose one way to quantify visual entropy using existing mathematical tools.

We then suggest that there is a natural semantic fit between an increasing scale of visual entropy of a shape and an increasing scale of uncertainty in data that can provide a novel way to define a set of glyphs for representing uncertainty in data visualizations, and test this hypothesis in a series of experiments.

\begin{figure}[htb]
  \centering
  \includegraphics[width=.95\linewidth]{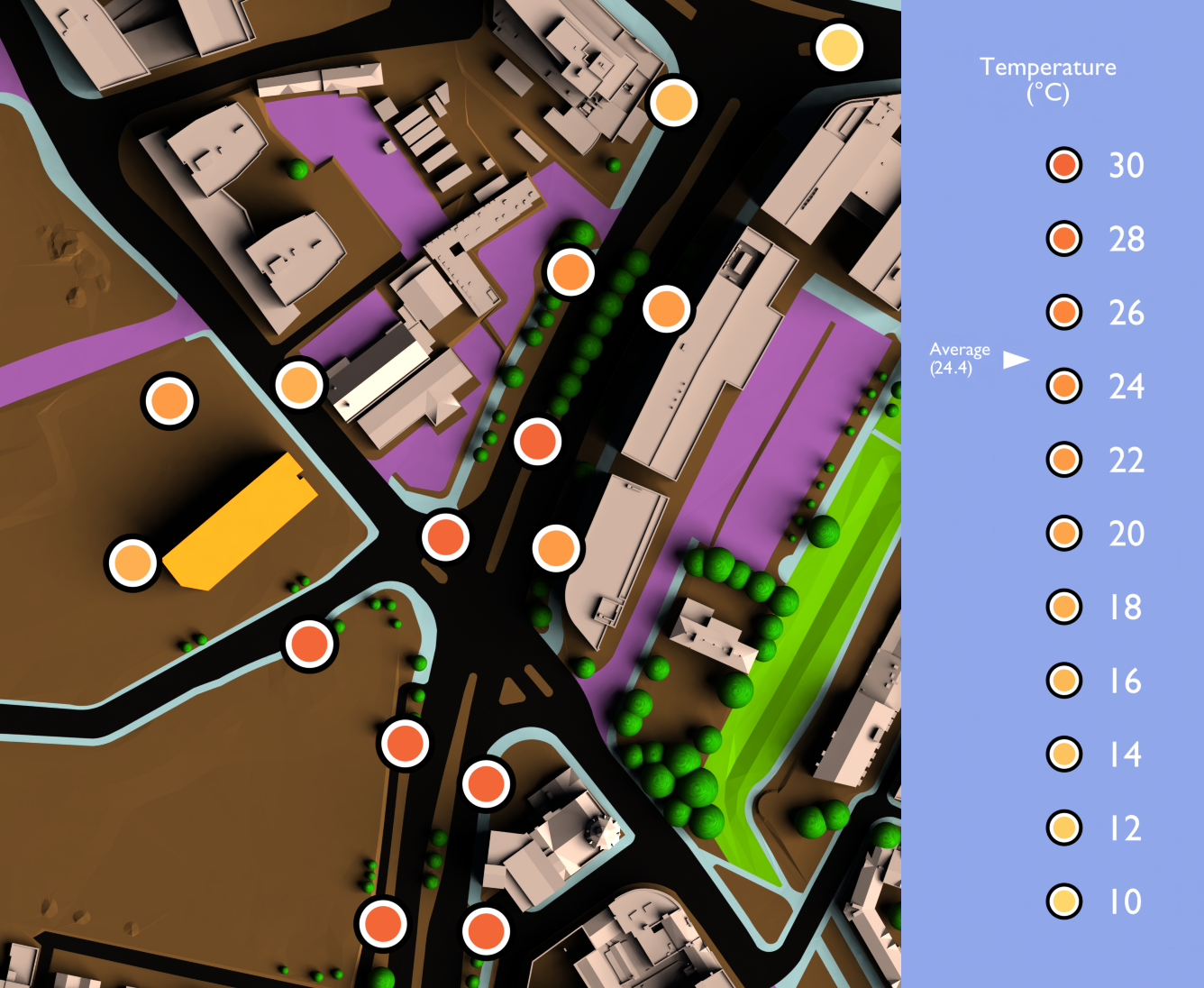}
  \caption{\label{figIoTSensors}
           Target style glyphs representing sensor values at locations in central Newcastle-upon-Tyne, the color represents the mean temperature from the last hour of readings for one sensor.}
\end{figure}

Specifically, we address three research questions:
\begin{enumerate}
\item Can we use visual entropy as a measure of shape complexity that predicts the human ranking of simple and complex shapes?
\item Can we use visual entropy to construct categorical and/or continuous scales of glyphs in visualizations?
\item Can we use glyphs defined on a scale of visual entropy in environmentally valid situations where representation of uncertainty is important for task success?
\end{enumerate}
We propose and test a novel set of glyphs for representing ordinal values and a novel application of these to the visualization of uncertainty in an environmental monitoring context using sensor networks. Our initial motivation for this work was the representation of urban sensor data from a network of Internet-Of-Things (IoT) sensors \cite{holliman2019Petascale} across Newcastle-upon-Tyne, a small subset of which are shown in Fig.~\ref{figIoTSensors}. 

In the featured data, each sensor reports a continuous stream of values for a range of measures including temperature, NO, humidity and others that are then summarised visually as an average over an hour. However, the quality of the sensors in the network is not consistent, it varies with both sensor cost and with time in the field. We wanted to find a simple but consistent way to represent the variance of each reading, as well as its mean, so that users could at a glance understand the level of uncertainty associated with each sensor’s readings. Other applications with similar uncertainty visualization needs include (but are not limited to) humanitarian or military operational planning, financial risk results and machine learning parameters.
\begin{figure*}
  \centering
  \includegraphics[width=.95\linewidth]{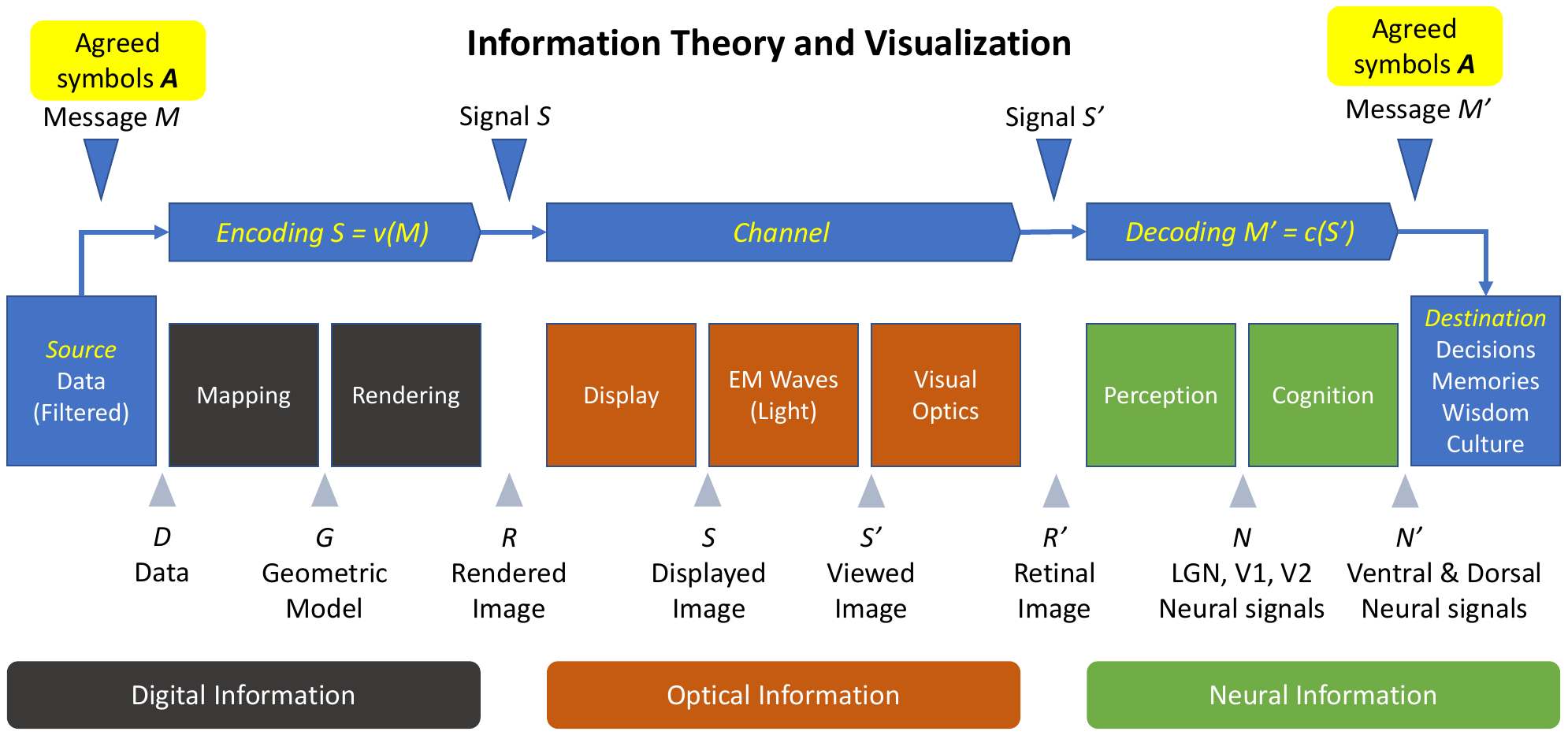}
  \caption{\label{figInfoTheory}
           The visualization pipeline mapped to an information theoretic communications pipeline as if the brain is an information theoretic receiver of visually coded messages M coded from an alphabet of symbols A, information is transformed physically twice as it flows along the pipeline from digital to optical and then from optical to neural signals.}
\end{figure*}

\section{Background}

\subsection{Information Theory and Visualization}
A number of articles have been published on the relationship between information theory and visualization \cite{vazquez2001viewpoint,rigau2007conceptualizing}. In Fig.~\ref{figInfoTheory} we illustrate how Shannon’s \cite{shannon1948mathematical} communication pipeline could map to the visualization pipeline. Encoding can be modelled as a process of image generation, communication as the optical path from display to retina and decoding as the process the brain uses to comprehend information encoded in the relayed image.
We tend to agree with \cite{maceachren2004maps} that information theory is a weak match as a model for the human (neural) part of this pipeline because the human brain does not act as an ideal decoder of visual codes in an information theoretic sense. Indeed, Shannon was explicit about this “semantic aspects of communication are irrelevant to the engineering problem” \cite{shannon1948mathematical}, and yet knowledge about perception and cognition is essential to the production of visualizations.
Recent work on understanding the link between the perceived and measured complexity of map visualizations suggests that psycho-physically informed models of visual complexity are able to predict subjective human rankings of complexity \cite{schnur2018measured}. This emphasizes the importance of perceptual factors as well as information theoretic factors in visual cognition.

\subsection{Entropy}
Information theory allows us to predict the average information content of a signal based on the cost of losslessly coding a message using the most efficient coding method possible. That is, it is an estimate of what is left after all possible redundancy has been removed from a signal, and in that sense is a direct measure of information content. This is helpful as it allows estimating channel capacity when designing communications networks \cite{cover1999elements}.
Entropy, $H(X)$,  of a random variable $X$, as given in (\ref{eqnshannonentropy}) is based on the probability $p(x_i )$ of any one symbol, $x_i$, in an alphabet, $A$ of size $n$, appearing in the message:
\begin{equation}
H(X)=\sum_{i=1}^{n} p(x_i) log_2\left( \frac{1}{p(x_i)} \right)
\label{eqnshannonentropy}
\end{equation}
Entropy tells us how many bits on average are needed to code any symbol from the alphabet in the signal. While this might appear to be a helpful way to optimise coding systems, it turns out that signals intended for human perception can be optimised to be perceptually lossless to a much greater degree. Examples of this include the JPEG \cite{pennebaker1992jpeg} and MPEG \cite{pereira2002mpeg} coding standards where great effort has been made to understand how much information can be omitted before errors becomes visible to the viewer of the image. Without the application of knowledge about the limits of human perception to lossy coding methods these standards and much visual internet traffic would be impractical or very significantly slower.

\subsection{Uncertainty}
Uncertainty by definition captures our \textit{lack of 
knowledge} about a value or outcome, most often expressed quantitatively, among the most widely used 
being Pearson’s standard deviation \cite{magnello2009karl} and Fisher’s 
variance \cite{fisher1919xv}.  
Uncertainty can be classified as aleatory or epistemic, depending on whether it is due to random variation or to unknown factors \cite{OHagan2004}. 
Semantically uncertainty is challenging, because a value that represents a lack of a property is not intuitively easy for non-experts and, at times experts to understand. 
One way to make the concept more accessible to humans is to rethink the language, i.e., when talking about uncertainty, it may be that a positive phrasing can be more helpful than negative, such as \textit{degree of certainty, level of certainty, confidence level, accuracy} and \textit{precision} \cite{hardman2009judgment}.
There is some agreed standardization in relation to levels of uncertainty, and there are national and international standards for reporting measurement uncertainty from e.g., metrology laboratories \cite{taylor1994guidelines}. In addition, some weather forecasts provide a degree of (un)certainty data, for example the UK Met office provides statements on precipitation in a standard form “There is a 70\% chance of rain.”, covering a defined time period \cite{MetOffice2022}. 

In this work, we aim to design and evaluate a clear visual solution to represent at least one type of uncertainty that does not rely on the viewer having statistical knowledge, but still conveys information about the uncertainty of a value which can be related back to the underlying statistical methods when needed. 
Perhaps the closest in concept to our aim here is the use of strings of asterisks categorical significance codes in conventional statistical reporting and software such as R \cite{RCoreTeam2018} for conveying the significance of model fitting outcomes where across a range of tests and model fitting methods the same set of significance codes are used. 
While it is broadly adopted and familiar to many people who are familiar with statistics conventions, one criticism we have of this asterisks representation is the choice to have a blank space as the symbol for significance in the interval $0.1 < p <= 1$. This conveys no information and it is not clear to the reader whether this is a test outcome or a printing error.

There is a long standing debate \cite{berkson1942tests,cairns2019doing} about over reliance on preset levels of significance and the related behaviors this generates in science. However, taking a broader view, we believe that it is valuable to try and engage a wide audience with visualizations that provide an everyday representation of measurements and predictions with varying levels of uncertainty and/or varying 
levels of significance. For researchers there already exists in-depth advice on the use and presentation of p-values, and other uncertainty measures, from professional bodies \cite{Wasserstein2016} and practical guidance on how to design experiments that severely test hypotheses\cite{cairns2019doing}.

\subsection{Approaches to Uncertainty Visualization}
Earlier surveys of uncertainty visualization (e.g., \cite{pang1997approaches}) identify seven methods which we categorize into one of two basic approaches: Those modifying the scene directly and those adding annotations to indicate levels of uncertainty. The approach that is most effective is clearly application dependent. 

Approaches to representing uncertainty visually often relate to summarizing the spread of values related to a measurement: dot plots, histograms, box plots \cite{tukey77}, confidence intervals and probability distribution functions \cite{brodlie2012review} provide ways to do this. These often presume some basic statistical knowledge on the part of the users, and an ability to interpret meaning from a spread of values. An empirical study of glyph-based approaches is presented by MacEachren et al. (2012) where authors show that the often-used visual channels of lightness and fuzziness perform well on their own \cite{maceachren2012visual}.

There are rigorous studies (e.g., \cite{bonneau2014overview}) highlighting that it is far from routine for visualizations to include uncertainty information even though it is fundamental to informed decision making. 
Contemporary workshops run by government agencies (e.g., \cite{Bowman2018}) have highlighted that even in critical operational planning situations there is a real difficulty in finding ways to convey uncertainty to high level decisions makers. 
It remains an open question how best to visualize uncertainty \cite{landesberger2017visualization,padilla2022}, particularly when a single glyph must represent both a variable’s mean value and its uncertainty. 

Communicating aleatoric and epistemic uncertainty is an active area of study and debate\cite{vanDerBles2019communicating}. In this study, we concentrate mainly on examples of aleatoric (scientific, measurable) uncertainty, but we also consider the issues of epistemic (unknown but knowable) uncertainty when 
we consider how to visually represent values we do not know. As a concrete example, for IoT sensors we can measure the variance of a sensor over an hour and visualize this variance as the sensor’s (aleatoric) uncertainty. 
However, in some cases only one, or no, measurements are available, and we have no information on the variance. In our concluding discussion we propose a distinctive representation for epistemic uncertainty i.e. of a value that we could in principle know but in practice do not. 

\section{Visual Entropy}

If we imagine the human brain to be an ideal Shannon decoder, then it should decode and respond to signals differently with differing levels of entropy.
In practice, though, the brain is much more than a decoder of signals, it generates its own hypotheses about the world, taking decisions based on partial information and weighting information in highly non-linear ways \cite{hardman2009judgment}. 
What also seems clear is that it does not need to produce a realistic decoding of the world around it \cite{svarverud2012demonstration}, it is instead very efficient in extracting and using just enough information to complete a task.

A consequence of the mind’s ability to hypothesize novel ideas and impute additional information is that it 
does not obey the data processing inequality \cite{cover1999elements}. It can and does add information at the end of the visual pipeline, this observation leads to the conclusion that it is far from an ideal information theoretic processor in the sense of Shannon’s theory. 

However, here we propose that even if the brain is a noisy decoder of signals it still is a decoder of signals and will have a sensitivity to differing levels of entropy in signals. We hypothesize that we should be able to use levels of entropy as a visual cue in coding information in visualizations, in the same way we already use varying color, brightness, size and other visual cues \cite{bertin1983semiology}. 

Following the definition of information entropy above, we define visual entropy $V(X)$  as the average cost of coding visual symbols $v_i$ from a visual alphabet which have a probability of appearing in the message of $p(v_i)$:

\begin{equation}
V(X)=\sum_{i=1}^{n} p(v_i) log_2\left( \frac{1}{p(v_i)} \right)
\label{eqnvisualentropy}
\end{equation}

The higher the visual entropy $V(X)$  the more information is contained in the visual message, the more visual complexity the message contains. This argument has close similarities to the definition of viewpoint entropy presented in \cite{vazquez2001viewpoint}, but we keep this theoretical definition more general, rather than incorporate specific concepts of cameras and polygons.  
The visual entropy of a message could also be viewed as a measure of how incompressible it is, in this sense a more complex visual message will need more coded information to be sent in the signal, it cannot be coded as a simple signal.

To transform this theoretical construct to a practically meaningful visualization cue, we will consider visual entropy as analogous to some extent to visual complexity. Visual signals with higher visual entropy have a higher visual information content, requiring more bits on average to be coded for lossless transmission. A smooth sine wave for example can be coded in fewer bits than a signal consisting of uniform white noise. Our use of the term complexity here relates to perceived visual complexity and signal incompressibility, rather than to the generation of complex phenomena from chance chaotic behavior.

To make practical use of visual entropy, we next discuss the design of visual glyphs that both represent data and its uncertainty. We consider how to practically measure visual entropy and propose an extended glyph design that uses visual entropy to represent uncertainty values. We then report an experiment testing our glyph designs to evaluate whether they can represent a scale of uncertainty. Finally, we test the glyphs in an environmentally valid application situation where we ask users to search for the most and least reliable sensors across a 3D map.

\section{Glyphs for Urban IoT Data}

\begin{figure}[htb]
  \centering
  \includegraphics[width=.9\linewidth]{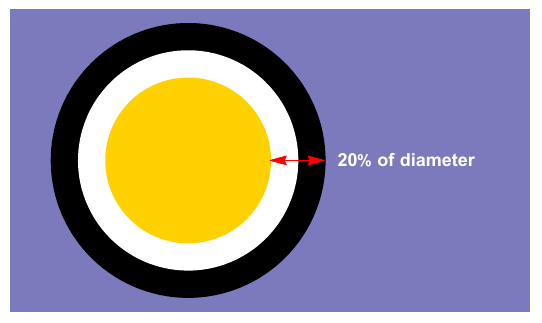}
  \caption{\label{figbasicGlyph}
           The glyph design we adopted to represent a measurement 
from an individual IoT sensor, the two outer rings are designed to 
have a width of 20\% of the total diameter, the colored central disc 
represent the sensors mean value on a predefined color scale. }
\end{figure}

In our previous research, we have implemented a number of glyph designs for representing urban environmental data in 3D city models \cite{holliman2017designing, holliman2019Petascale}. We currently prefer the glyph design in Fig.~\ref{figbasicGlyph}, as used in Fig.~\ref{figIoTSensors}, that, while located in a relevant position in 3D space, is presented to the viewer as a primarily 2D shape.

The design, shown in Fig.~\ref{figbasicGlyph}, took some aspects from that of a target of concentric rings and some from the design of the Landolt C optotypes \cite{landolt1899nouveaux}. The rings, a dark outer shape and a light inner shape, were chosen in order to highlight the glyph against both light and dark back-grounds. It also provides a level of self-contrast for the glyph. The total width of both outer rings is set to be 20\% of the diameter of the whole target, matching that aspect of the Landolt optotype design.

We use the central disc to represent data value, and typically we use color to do this, following color scale standards set in the literature. In the examples here, the data value represented in the central disk is temperature and adopts the colors used by the UK Met Office \cite{MetOffice2022}. In our visualizations, a color legend is usually displayed on or near the visualization. 

The visualizations presented here use physically-based path tracing for the graphical rendering stage implemented with Blender Cycles \cite{blender2018}. Our goal is to use realistic lighting simulation to help engage viewers in the 3D image however we therefore need to make sure this realistic lighting does not alter the glyphs information carrying appearance. We do this using a number of techniques including flat shading colors and rotating glyphs to face the camera, full details of the glyph implementation are given in the Appendix.

\section{Entropy as a Visual Cue}

A question that was raised when presenting our urban data visualizations was \textit{how much is it possible to rely on the sensor data?} Expert members of an audience will be aware that different sensors can have very different accuracy and precision related in part to their cost. To help answer this question in a visual form, we started to consider how we could represent the uncertainty of measurements at the same time as the value of measurements in our glyphs.

Looking for a visual cue supported by existing scientific evidence that we could use in parallel to color to represent uncertainty, we considered shape as a possible cue \cite{wolfe1998attention} and reviewed a study \cite{cutting1987fractal} on the human perception of fractal shape. Cutting and Garvin (1987) \cite{cutting1987fractal} demonstrated that certain fractal generation parameters correlated well with perceptual ordering of perceived shape complexity. More recent work in visual search suggests the cues of size and frequency, which help form shape differences, are both reported to guide (direct) attention in visual search \cite{wolfe2017five}. This led us to hypothesise whether varying the levels of visual entropy, as a measure of shape complexity, might be used as a visual cue, and ultimately as a scale to represent ordinal categorical or interval numerical values.

To implement and evaluate this hypothesis, we need two things: a practical measure of visual entropy and a geometric representation for the glyphs that as it varies exhibits varying levels of visual entropy.

\begin{figure}[htb]
  \centering
  \includegraphics[width=.95\linewidth]{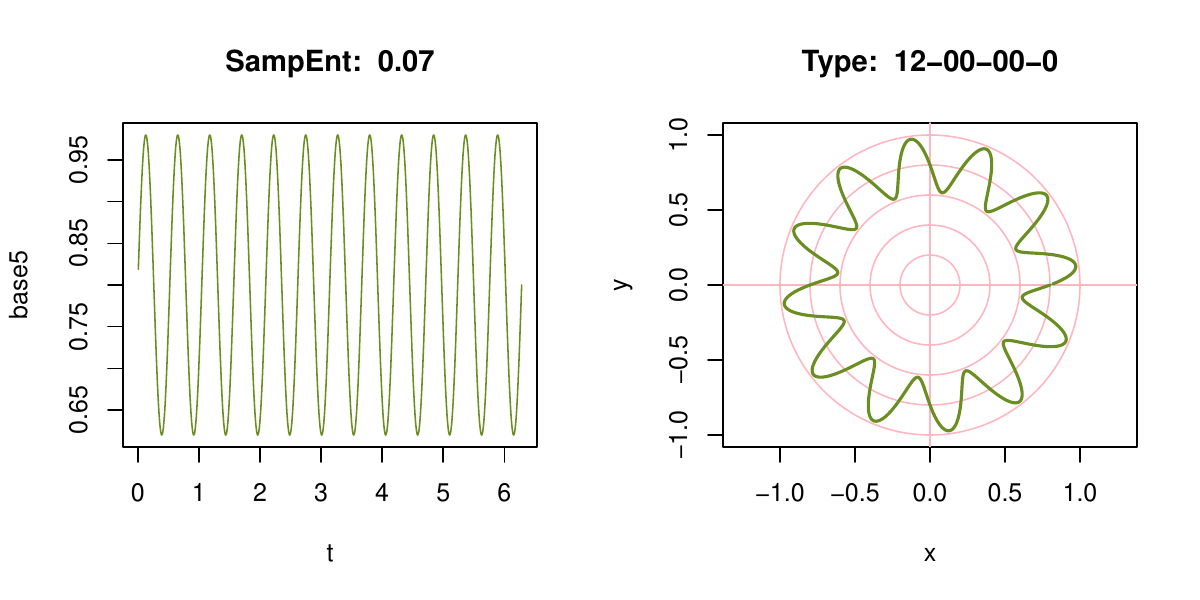}
  \caption{\label{fig.entropy}
           The message M on the left is coded as the signal S on the right, by plotting the message in 360 degrees on a polar plot; we estimate the visual entropy of the glyph as the sample entropy of the message. }
\end{figure}

We start by considering how a visual signal $S=v(M)$ can be generated by a coding function $v$  from an abstract message $M$. The message we encode need have no direct meaning, but for our purposes it does need to be able to represent variable levels of entropy. The signal is the geometric representation of the message that we will eventually render in our glyphs. This first step in the glyph generation is illustrated in Fig.~\ref{fig.entropy}.

To estimate the visual entropy of the message, we calculate the sample entropy \cite{richman2004sample} of the message before it is coded as a geometric shape. Sample entropy, and the related function approximate entropy, provide an estimate for the regularity and unpredictability of a data series, they are often used for comparing time series such as electro-cardiograms. Sample entropy has less dependence on the series length and can be more consistent across series \cite{yentes2013appropriate}, hence we adopt it for our calculations. We use the sample entropy implementation in the R pracma package \cite{Borchers} with parameter settings of n=2,r=0.2 based on guidance from \cite{yentes2013appropriate} to estimate the visual entropy in our glyphs.

\begin{figure}[htb]
  \centering
  \includegraphics[width=.45\linewidth]{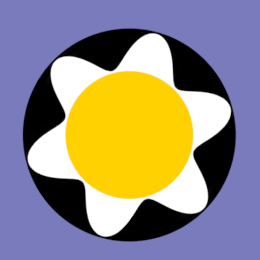}\hfill
  \includegraphics[width=.45\linewidth]{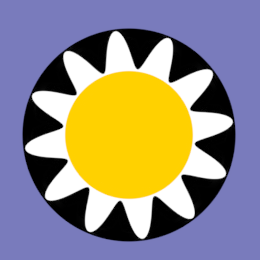}
  \caption{\label{fig.glyph}
           Two glyphs created in Blender from signals with different sample entropy measures, the central temperature color mapping is the same in each glyph, we hypothesize the visual entropy of the surrounding shape can represent an orthogonal value such as uncertainty, or variance of the temperature value.}
\end{figure}

This gives us a route to create geometric shapes with measurably varying levels of visual entropy. To add these to our existing glyphs in Blender we export the signal shape from R and import it to Blender as an extruded polygon. These polygons are then used to replace the inner white disc in the glyph, as illustrated in Fig.~\ref{fig.glyph}.

The resulting designs (Fig.~\ref{fig.glyph}) provide an ability to create glyphs representing a value, such as mean temperature using color in the central disc and a second value, such as uncertainty in the variation of the surrounding shape. We now need to evaluate whether viewers can naturally order glyphs of differing visual entropy. This will begin to demonstrate whether we have a categorical ordinal (or potentially numerical interval) scale of visual entropy available in these glyph designs. 

\section{Representing the Null Case}

An important consideration is the case where there is missing uncertainty data. While the glyphs in Fig.~\ref{fig.glyph} can represent a range of uncertainty, we also need a glyph design that represents the case when we have no data on uncertainty. To clarify, this is for the case where we have a data value (mean temperature) to present, we just have no information on its associated uncertainty (variance) level.

\begin{figure}[htb]
  \centering
  \includegraphics[width=.45\linewidth]{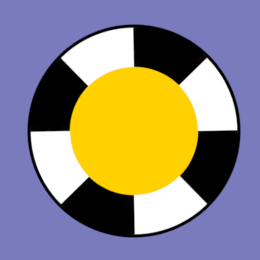}
  \caption{\label{fig.nullglyph}
           A possible glyph design for the NULL case (epistemic or ontological uncertainty), where we have no uncertainty measure and/or no data value. }
\end{figure}

To implement the null case, we considered various warning symbols searching for a symbol that is visually different in design to those in Fig. 5. Our current proposal is the glyph in Fig.~\ref{fig.nullglyph}, which is significantly different to our other designs and is related to familiar warning symbols.

\section{Evaluating the Visual Entropy Glyphs}

\begin{figure*}
  \centering
  \includegraphics[width=.95\linewidth]{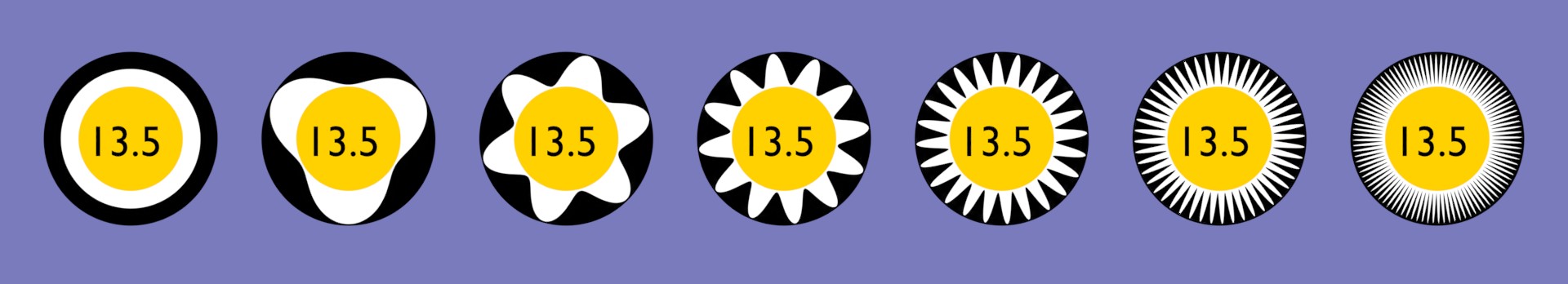}
  \caption{\label{fig.allglyphs}
          The set of glyphs used in the user evaluation, ordered by the calculated visual entropy value of the generating message, as shown below each glyph, when viewed on the controlled test displays the glyph on the far right is displayed at or above the limit of human spatial acuity (10 cpd), in the results these images are referred to by the labels A to G from left to right.}
\end{figure*}

Based on pilot testing in our laboratory in Newcastle (UK), we choose to evaluate the glyphs shown in Fig. \ref{fig.allglyphs}.
This set of glyphs uses a single sine wave as the generating message, at varying frequencies. These are similar to radial frequency patterns which have been shown to be detectable at low amplitudes in 
the psychophysics literature \cite{wilkinson1998detection}. Recent studies support that these are discriminable shapes based on frequency differences \cite{dickinson2018visual}, are identifiable even if only the shape convexities are visible \cite{schmidtmann2015shape} and can represent numerical order \cite{chung2016ordered}. In addition, in these trials we added a numerical indication of the temperature value on the glyphs, the same value of 13.5 degrees Celsius was used in all trials. 

The increase in sine wave frequency from glyph to glyph was set on a geometric scale, the frequency doubling for each new additional glyph. 
This is similar to the logarithmic increase between levels on a logMAR visual acuity chart \cite{jackson2004visual}, but here the change increases at a significantly greater rate per level (2x rather than 1.26x) in part 
so that the full range of visual acuity is used in a smaller number of glyphs, and in part because this evaluation is designed to determine whether an ordering of visual entropy exists rather than to determine the level of just noticeable difference (JND) between glyphs. 
This anticipated that, like many aspects of human perception \cite{fechner1966elements,varshney2013we}, there would be a logarithmic response to the visual entropy stimulus.

\subsection{Experimental Method and Apparatus}
The experiment to test the visual entropy glyphs has two objectives. First to test the hypothesis that there is a perceived rank ordering between the glyphs shown in  Fig. \ref{fig.allglyphs} Second to demonstrate whether this order is predicted by our numerical measure of visual entropy. 

To evaluate whether there is a rank ordering between the glyphs a two alternative forced choice (2AFC) method \cite{hautus2021detection} of glyph image pair comparisons was implemented where all paired permutations of the set of glyphs are shown to each participant, except for those where the pairs would be the same shape. The experiment was replicated in different countries, in three different environments, two physical-in-person (Perth, Australia and Brugg-Windisch, Switzerland), and in addition, online using Mechanical Turk.

 The stimulus presentation was implemented using the PsychoPy toolbox \cite{peirce2007psychopy} and answers for each pair comparison, a left or right arrow key-press, were recorded in addition to the time taken to enter the answer. Each participant saw a different random order of pairs, determined by PsychoPy’s random number generator. The stimuli for the in-person trials were presented at a viewing distance of approximately 500mm for all participants, differences between the environments are discussed below.

The Perth trials were presented on a 11.8” FHD LCD monitor built into a Lenovo laptop, the Swiss trials were presented on a 15.6" FHD LCD monitor built into a HP laptop, and the Mechanical Turk (MTurk) trials were presented remotely on unconstrained display devices. As a result of the unconstrained environment, we anticipated some variability in the outcomes of the MTurk trials and designed a set of data cleaning methods to detect outliers described below alongside the results.

The instructions that each participant read before the trial were:
{\small
\begin{verbatim}
You will see a series of image pairs.
Each image represents a value and also 
represents a level of uncertainty.

More complex shapes represent more uncertainty.

Choose which image represents the most 
uncertain value to you.

Left arrow for left. Right arrow for right.
Press space when ready.
\end{verbatim}
}

Note, we considered the use of the word complex with some care, as it was clear that visual entropy would not be a widely understood description of shape differences in the images. Participants then began the trial where they were presented with all 42 pairwise permutations of the images in Fig. \ref{fig.allglyphs}, this included reversed order image pairs.

\subsection{Ethical Approvals and Participants}
Ethics approval was granted at Newcastle University, UK to run a global experiment (inside and outside the EU) and was also independently requested at Curtin University, Perth, Australia. Participants in all the trials gave consent for the data from the trials to be used and communicated worldwide by the investigators for the purposes of the study. We requested very limited personal data, whether participants had normal or corrected to normal visual acuity and whether they were aware of any color deficiency in their visual ability. While colour deficiency was unlikely effect the results in a shape complexity comparison experiment, we recorded this in case it had a bearing on the saliency of the central colored disc.

In the Perth trial nineteen participants, $n=17$ excluding outliers, were drawn from staff and students at the Curtin Institute for Computation and the Curtin High Impact Visualization Environment. All reported normal or corrected to normal vision. One participant reported a color deficiency but judging color was not a required part of this experiment, therefore their results were included in the analysis. 

In the Swiss trial, twenty-two participants, $n=20$ excluding outliers, were drawn from staff and students at the University of Applied Sciences and Arts Northwestern (FHNW) in Brugg-Windisch, Switzerland. All reported normal or corrected to normal vision.

In the online MTurk trial, fifty-six participants, $n=50$ excluding outliers, were drawn from the available pool on Amazon Mechanical Turk. All reported normal or corrected to normal vision.

%There was no plan to analyse sub-groups and therefore no justification to record factors such as gender, age and previous expertise in this set of trials. 
Our primary intent is to establish whether there is a broad response across all viewers to the glyphs in line with our hypotheses.

\subsection{Omnibus Testing}
Before evaluating the experimental hypothesis we apply a series of omnibus G-tests to check whether there are variations from our expected outcomes at the different locations.
These omnibus tests allow us to understand; 1) if there is any variation from the expected proportions of correct answers within the glyph results at each location, 2) whether there is any significant variation across the replications at the three locations, 3) whether there is any unexpected variation in the data pooled across all three samples (AU, CH, MTurk) and finally 4) to use the additive property of the G-Test to check that overall the data fits our predictions. The G-test (compared to chi-squared and exact tests) is recommended for situations with larger numbers of observations \cite{StatsTest2022} and when replicating studies using multiple location testing \cite{McDonald2022}.

\subsubsection{G-Test of Goodness of Fit by Location}
Our first omnibus null hypothesis is that at each individual location the proportion of correct answers for each of the glyphs is the same, i.e. $1/7$ of the total number of trials. This is a reasonable null hypothesis since we designed the glyphs to be above the acuity threshold and to be clearly distinguishable on an exponential scale of increasing complexity. The alternative hypothesis is that there is a statistically significant difference in the proportions of correct answers for some of the glyphs at one or more locations, and if so this will need further analysis.

Using a G-Test for Goodness of Fit at each location we find the following; for Perth $(G=1.5197, p=0.9582, df=6)$, for Swiss $(G=0.6137, p=0.9962, df=6)$ and for MTurk $(G=1.5289, p=0.9576, df=6)$. In each location there is no evidence (since all $p >> \alpha=0.05$ ) for rejecting the null hypothesis and therefore no evidence that there is any unexpected variation in the outcomes at each of  three locations.

\subsubsection{G-Test of Independence}
The second omnibus null hypothesis tested is that the relative proportions of correct answers are the same across the three different locations. If the relative proportions are not the same we will need to analyse locations separately and will be unable to pool them.

This hypothesis is evaluated using a G-Test of Independence with a $3x7$ contingency table (3 locations, 7 glyphs). The result is $(G=1.1246, p=0.99997, df=12)$. There is no evidence (since $p >> \alpha=0.05$) for rejecting the null hypothesis and we can conclude we can now consider the pooled results.

\subsubsection{G-Test for Goodness of Fit for Pooled Results}
The third omnibus null hypothesis is again the equal probability hypothesis for each glyph but now we apply this for the pooled results combined across all three locations (n=87).

Using a G-Test for Goodness of Fit for the pooled results we find the result  
$(G=2.5378, p=0.8642, df=6)$ and again we cannot reject the null hypothesis (since $p >> \alpha=0.05$) and conclude that there is no evidence for any unexplained variation between the proportions of correct answers for the glyphs in the pooled results.

\subsubsection{Outcome by total G-value}
The final omnibus test checks the overall outcome is not inconsistent with our null hypothesis of equal probability. We use the additive property of the individual location G-Tests from step {\it 1)}, sum these to give the total G-value $(G=3.6623, df=18)$, looking this up in a chi-square table gives a p-value $(p=0.99988)$. Again we cannot reject the null hypothesis (since $p >> \alpha=0.05$) and we conclude that overall there is no evidence of any unexpected variation in the overall outcome across glyphs and locations.

In summary, over all four omnibus tests we have found no statistically significant evidence to reject the null hypothesis, that the glyphs have equal probability of being correctly chosen, and we have found no evidence to suggest there are statistically significant differences between the replication locations. We can therefore move on to consider the pooled data as a whole in testing the main experimental hypothesis.

\subsection{Analysis of the Pooled Results}
The design goal for the visual entropy glyphs is to create a perceived order among the glyphs that can be predicted by the visual entropy calculation for the enclosing shape. 

We use the pooled response data for each glyph, shown in Table~\ref{table.results01}, to test the hypothesis that participants agree with the predicted order. 
The independent variable is each glyph type $(A,B,C,D,E,F,G)$ and the dependent variable is the proportion of correct choices made about the order of each glyph in the pairwise comparisons with all six other glyphs.
We deem the response "correct" when a participant's choice is as predicted by the visual entropy calculation. With $n=87$ participants the total number of trials per glyph is $87*6 = 522$, and this is therefore the maximum correct score each glyph could achieve.

\begin{table}
\centering
\caption{Pooled results for the glyph pairwise order comparisons with the outcomes of the exact binomial test for each glyph.}

\begin{center}  
\includegraphics[width=0.48\textwidth]{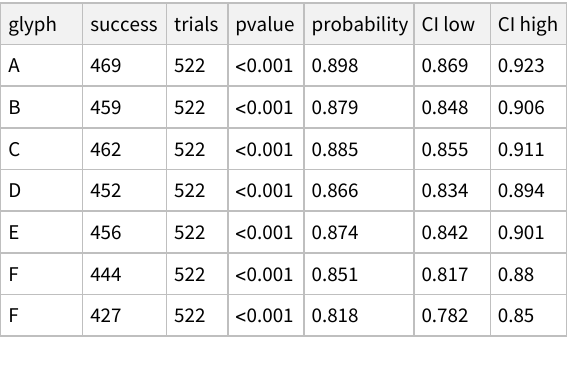} 
\end{center}     

\label{table.results01}

\end{table}

The null hypothesis, $H0$, for this analysis is that participants will do no better than chance at choosing between glyphs in the order predicted by the visual entropy calculations ($probability <= 0.5$ ) . The alternative hypothesis, $Ha$, is that a majority of participants do agree with the predicted order ($probability > 0.5$). 

\begin{figure}[ht]
  \centering
  \includegraphics[width=.95 \linewidth]{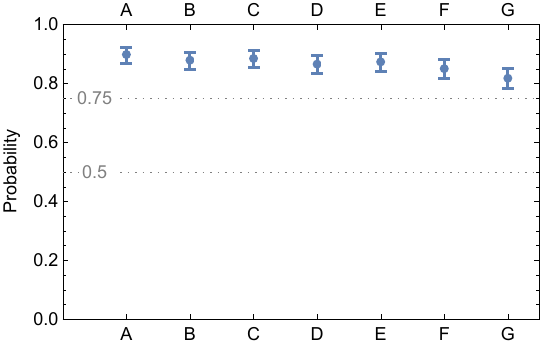}
  \caption{\label{fig.glyphorder}
           Pooled results showing 95\% confidence intervals, the null hypothesis, $H0$, is the ${probability} < 0.5$ that the majority of participants in agreement with our entropy prediction is no better than chance. }
\end{figure}

As the results are a count of categorical choices, we apply an exact binomial test against a simple majority for each glyph. Because the test is repeated for each level (glyph) we apply a Bonferroni correction to the alpha threshold for the p-values and use $\alpha=0.05/7=0.007$ as the significance level. The results of the binomial tests are given in Table~\ref{table.results01} and illustrated in Fig.~\ref{fig.glyphorder}. 

For all seven glyphs tested, the p-values are $< 0.001$ and as these are all smaller than $\alpha$ we can reject the null hypothesis of chance performance, and accept the experimental hypothesis that a majority of participants' choices agree with the predicted order.

\subsection{Effect Size}

One approach for estimating effect size for binomial tests, recommended in \cite{peter2022}, is to calculate Cohen's $g$ \cite{cohen1988}.
This method can be used to estimate the effect size for probabilities only in comparison to a chance result of 50\% correct answers. 
Cohen's $g$ effect sizes are calculated for each of the glyphs as shown in Table~\ref{table.effectsize} these results suggest that the effect size for all the glyphs is Large ($g>0.25$).

As discussed in detail in \cite{cairns2019doing} the practical significance of effect size depends on context.
Here, given the high probability of users choosing the correct order, it seems reasonable for us to conclude the practical effect size will be strong. 
That is, very often users will judge the order of the glyphs as predicted, and this may be even more often if in practice we display a legend for the user to refer to. 
Equally, we need to be mindful, as this does not mean all users will always make the predicted judgement about the glyphs and in safety critical situations further empirical testing would be advisable to be assured of the required level of practical effect size. 

\begin{table}[h]
  \centering
    \caption{\label{table.effectsize}
           Effect size estimate using Cohen's $g$ for each of the glyph binomial tests, the inset to the right is the interpretation of $g$ given by Cohen.}
  \includegraphics[width=.95 \linewidth]{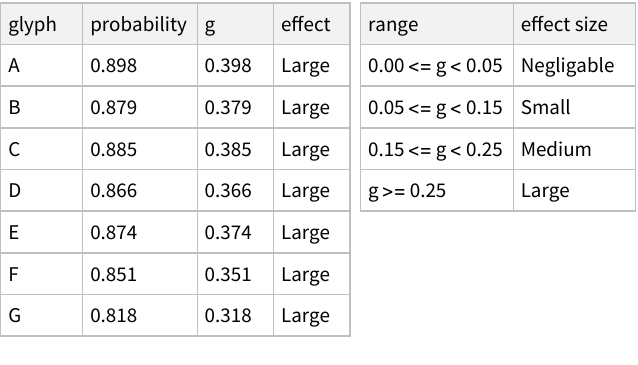}
\end{table}

\subsection{Response Times (RT)}

To evaluate the seven glyphs at the three different global locations, we had to allow the test environment to vary between the Perth, Switzerland and the online experiments. We anticipate this could introduce either random or systematic noise into the results. On the other hand, this is an ecologically valid testing context since in real-world visualization applications, the viewing environment will differ between users. We found no evidence for a statistically significant variation in accuracy (probability of choosing the predicted response) due to location or glyph type in the pairwise glyph order experiments reported above, and we now test the response times.

To select the appropriate omnibus tests, we tested for normality in the RT data using the Shapiro-Wilk method, across all locations this suggested $(W=0.5578, p < 0.001)$ we had to reject the null hypothesis and accept that the data is not normally distributed. This result also holds if we test the data separately at each location. Therefore we needed a non-parametric omnibus test and, as there is no direct equivalent for the two-way ANOVA for this experimental design, we used two separate one-way Kruskal-Wallis tests for the two independent variables location and glyphs.

\subsubsection{RT omnibus testing}

We considered the RT variation between the glyphs (seven levels) first using a one-way Kruskal-Wallis test. 
The result is $(\chi^2= 0.24288, {df} = 6, p = 0.9997)$ and as a result we cannot reject the null hypothesis $H0$ that the difference in the means of the RT between the glyphs is simply due to chance, with $p >> \alpha = 0.005$

We use the same type of non-parametric test to consider the overall effect of location (three levels) on the response times with the result $(\chi^2= 79.428, df = 2, p < 0.001 )$. In this case we do reject the null hypothesis and accept the alternative hypothesis that there seems to be statistically significant variation between the response times in each location beyond that expected by chance.

\begin{table}[h]
  \centering
    \caption{\label{table.rtdunn}
           Post-hoc Dunn test results comparing the response time (RT) between all location pairs, in all cases the null hypothesis is rejected suggesting the mean RT differ statistically significantly between all locations.}
  \includegraphics[width=.65 \linewidth]{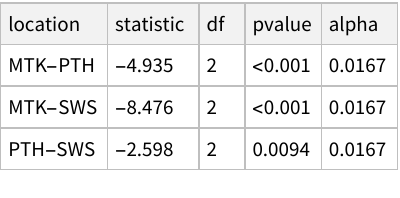}

\end{table}
To understand more about the variation between locations we use the Dunn post-hoc test against the test criterion $\alpha = 0.05 / 3 = 0.01667$, results are shown in Table~\ref{table.rtdunn}.
In all three possible location pair comparisons the null hypothesis can be rejected suggesting the RT differ statistically significantly between each of the locations.

As noted above we anticipated some variation between locations in at least some of the results due to test environment differences, and this analysis does suggest that some uncontrolled factors, such as varying screen size (phone or laptop/desktop), response method (mouse click or touch), or the experimental context (in person vs remote) are effecting the response times by location.

\section{Application Domain Testing}

Our results above suggest that the participants perceive the visual entropy glyphs in an order that is in agreement with our visual entropy predictions. We should therefore be able to use them to represent ordered categorical information or quantized numerical data on interval or ratio scales. Thus, in this section we examine the use of visual entropy glyphs in a specific context. Our urban digital twin application, see Fig.~\ref{figIoTSensors}, has a requirement that we display both a sensor’s mean value and its variance, so that end users, for example policy makers, can see at a glance which sensors they can rely on most.

\begin{figure}[htb]
  \centering
  \includegraphics[width=.95\linewidth]{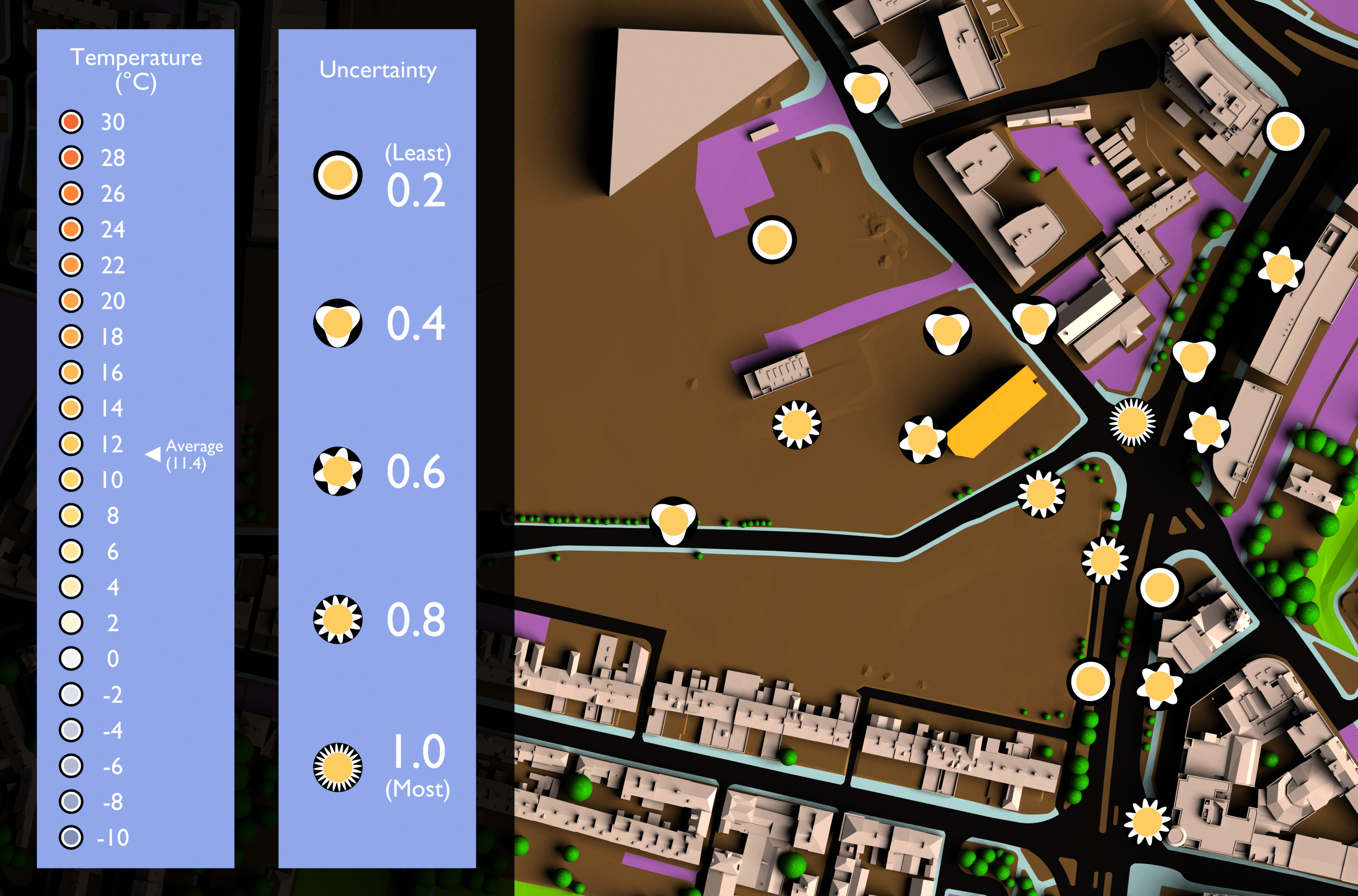}
  \caption{\label{fig.appTest}
           The urban temperature data visualization showing both hourly mean temperature values using the MetOffice color scale and the variance of those values using our new visual entropy scale, this image is an example of the high uncertainty target-present stimulus used in the experiment described below.}
\end{figure}

Previously, we displayed sensor data as the mean value over an hour using the MetOffice color scale for temperature, here we now also visualize uncertainty as the variance calculated over the same hour using visual entropy glyphs, see Fig.~\ref{fig.appTest}.
To set the range of the uncertainty scale, we need to calculate the range of variance for sensors in view, potentially over the whole city, so that we can calculate the minimum and maximum values on this scale.

\subsection{Experimental Method and Apparatus}

To evaluate whether this representation can be effective, we designed a signal detection experiment that requires the participants to search for a glyph based on its level of uncertainty. This was a target-present/target-absent visual search for either a low uncertainty target or a high uncertainty target. A total of fifteen participants (n=15), students and staff at Newcastle University (UK), each viewed forty images (therefore amounting to a total of 600 trials), searching for the least uncertainty glyph in ten target-present and ten target-absent images and the same again for the highest uncertainty glyph.

The display used for this experiment was a Microsoft Surface Pro 4, a display with 2736x1824 0.094mm square pixels at a nominal viewing distance of 500mm. Given this geometry, we calculated that the 24-cycle glyph was at the 10 cpd limit of human vision for this display, therefore we selected the set of five glyphs shown in Fig.~\ref{fig.appTest}. Once again, we used PsychoPy to present the stimulus to the participants. Ethics approval was granted, details of which were given earlier.

Our hypothesis in this study is that the target-present glyphs should be easy to find because of the choice of log-scale increments in the generating frequency and as a result the discriminability should be high. If this is case, we also hypothesize that there should be a response time difference between target-present and target-absent trials.

\subsection{Results of the Domain Testing}

\begin{figure}[htb]
  \centering
  \includegraphics[width=.95\linewidth]{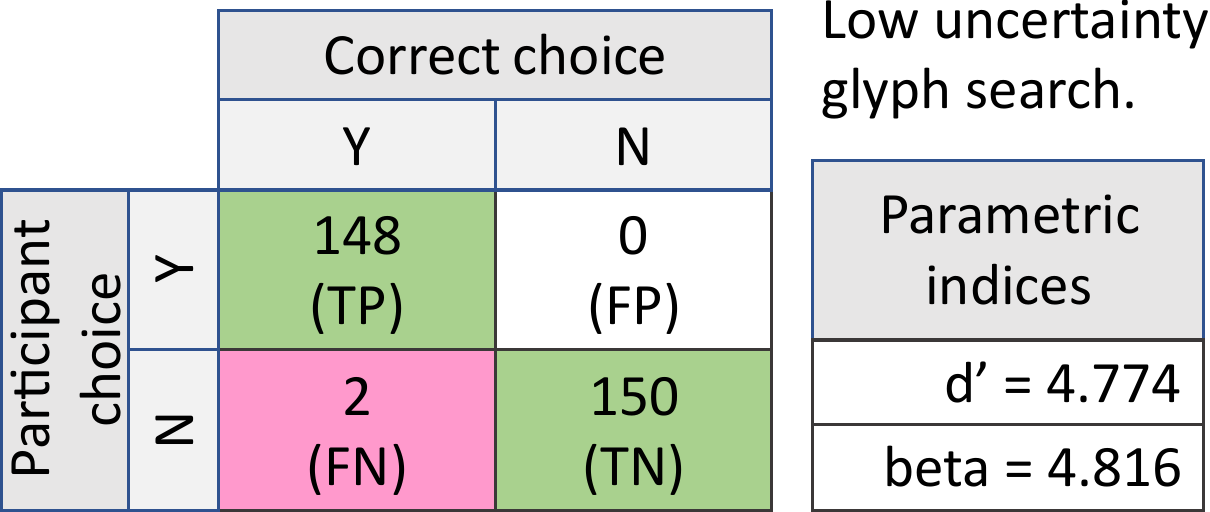}
  \caption{\label{fig.lowUConfusion}
          Confusion matrix for the low uncertainty glyph visual search from 150 target-present and 150 target-absent trials.}
\end{figure}

We analyzed the outcome data in R using the psycho package \cite{makowski2018psycho} from a total of 300 responses per glyph type (highest and lowest uncertainty), of which 150 were target-present and 150 target-absent in each group. The confusion matrix for the low uncertainty glyph searches is shown in Fig.~\ref{fig.lowUConfusion}. As hypothesized, the low uncertainty glyph is easy to find in visual search tasks, with high discriminability, d’, while the response bias, $\beta$ is towards answering (n) target-absent.

\begin{figure}
  \centering
  \includegraphics[width=.95\linewidth]{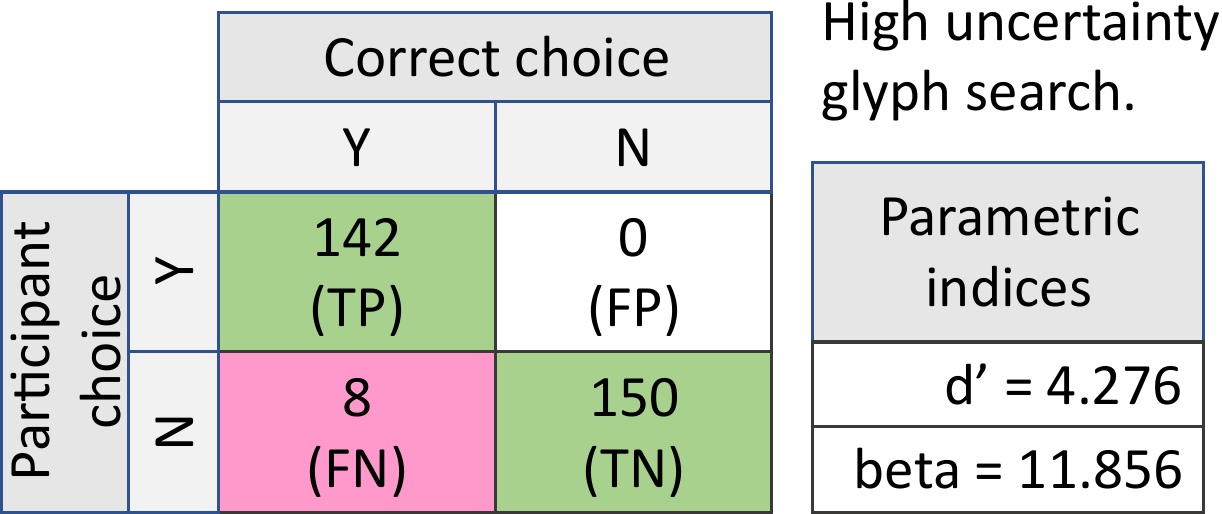}
  \caption{\label{fig.highUConfusion}
          Confusion matrix for the high uncertainty glyph visual search from 150 target-present and 150 target-absent trials.}
\end{figure}

The confusion matrix for the high uncertainty glyph in Fig. ~\ref{fig.highUConfusion} gives similar results, with a slightly higher response bias towards target-absent.

Response times were analyzed using two-way within-subjects t-tests, the results are shown in Table~\ref{table.rtresults}. As was hypothesized, target-absent trials took significantly longer to complete (approximately twice as long) than target-present trials.

There was no statistically significant difference between mean glyph search times for low and high uncertainty glyphs in the target-absent condition, and nor did we expect one as the search task is essentially the same.

There was a statistically significant difference $(p < 0.05)$ between target-present search times for the low- and high- uncertainty glyphs, with it taking on average 0.3 seconds longer to find the more complex, high uncertainty glyph. This was also supported by comments from some participants who reported the more complex glyph was harder to search for.

\begin{table}
  \centering
    \caption{\label{table.rtresults}
          Signal detection experiment mean response times, analysis suggests all differences are statistically significant ($p<0.0125$ with Bonferroni correction) except between glyph absent conditions.}
  \includegraphics[width=.95\linewidth]{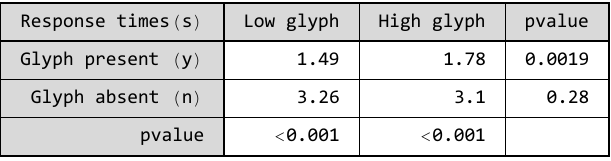}

\end{table}

In summary, the results from the application domain testing seem in agreement with our general hypotheses from the earlier sections: Participants could, with low error rates, search for and successfully find glyphs with different levels of uncertainty. We also identified the possibility that glyphs with higher visual entropy (more complex shapes) are slower to search for than those with low visual entropy.

\section{Discussion}

Uncertainty whether described as variance, probability, confidence or one of many other measures, can be difficult to visualize for experts and non-experts. We have considered whether a new form of glyph, derived from a theoretical consideration of visual entropy and perceptual results on visual complexity, could more clearly represent uncertainty measures on an ordered scale. We validated our proposed scale of visual entropy first in a ranking experiment, and presented a first evaluation that the visual entropy glyphs can be effective in specific contexts, e.g., in our digital twins application domain linked to urban monitoring and smart cities.

There is similarity between our work and the ideas recently described in \cite{gortler2017} where an informal argument for the use of wavelength and amplitude for representing uncertainty is made. Here we have provided a theoretical basis for this approach, a rigorous evaluation of glyph ranking and an application domain validation of the effectiveness of visual entropy glyphs.

Our starting point was to consider the visualization pipeline and its representation using information theoretic concepts, while this is a useful approach to some extent, Shannon’s information theory was not intended to model humans as information receivers. Humans in many ways are noisy decoders of information signals creating hypotheses and imputing data based on priors from many sources. Nonetheless, we argue that humans should be sensitive to underlying entropy levels in a signal and that visual entropy could therefore be used as a visual scale to represent values, e.g. as we investigated uncertainty.

To test this in practice, we designed a novel set of glyphs, similar to radial frequency patterns used in vision science, which we hypothesised would have increasing levels of visual entropy predicted by a calculation of Sample Entropy. We found from a user trial (pooled n=87) run in three locations evaluating pairwise glyph comparisons that participants ranked a set of glyphs in the order predicted by Sample Entropy. While this is the case for this set of glyphs, there remain several open questions:
\begin{itemize}
\item	How well does sample entropy predict human ordering of glyph shapes more generally?
\item	Are there better predictors of visual entropy available, or that could be developed?
\item	What is the just noticeable difference (JND) between glyphs of this design?
\end{itemize}

To test the glyphs in an application domain, we ran a signal detection experiment requiring participants to search for the lowest and highest uncertainty glyphs. This demonstrated that users could locate visual entropy glyphs when they were associated with levels of uncertainty. However, search time increased with increased visual entropy of the glyph by approximately $20\%$ and this suggests further open questions:
\begin{itemize}
\item	Are search times related in general to the visual entropy of a glyph?
\item	Does the visual entropy contrast between foreground glyph and background map vary search time?
\item	Are there ways to optimise search time independently of the glyph’s visual entropy?
\end{itemize}

In order to make best use of any glyph relying on shape features we need to take account of the size of the smallest distinguishable feature on the screen being used, usually measured in cycles per degree (cpd). For the visual entropy glyphs, we needed to know the on-screen pixel size of the glyph, the displays' dimensions and physical pixel size and the users viewing distance. These are all relatively easy parameters to know, measure or estimate for common displays and we would recommend become a routine calculation for visualization systems. As we demonstrated when displayed outside a closed test environment visualization performance can vary, although in our case this only affected response time and not accuracy.

\section{Conclusion}
There are many visual representations of uncertainty in technical publications that work for statisticians and for technical audiences. It is harder to find good visual representations of uncertainty in the everyday media and in documents intended for non-technical high-level decision makers.

We set out a formal argument for the use of visual entropy as a visual coding scale for visually transmitted information. We hypothesised that even though the human brain is not an ideal information theoretic signal receiver, it should still be sensitive to varying levels of entropy in signals. Intuitively, we can consider entropy in this case to be analogous to visual complexity.
We then set out to rigorously evaluate this approach by creating a set of glyphs and using those glyphs to represent uncertainty. We addressed three research questions:

\begin{itemize}
\item Can we use visual entropy as a measure of shape complexity that predicts the human ranking of simple and complex shapes?
We demonstrated we were able to predict human ranking of glyphs, using sample entropy as a proxy for visual entropy, with high confidence.
\item Can we use visual entropy to construct categorical and/or continuous scales of glyphs in visualizations?
We demonstrated a natural ranking order among our proposed visual entropy glyphs allowing us to represent ordinal categorial, or numerical interval, data on a discrete scale.
\item Can we use glyphs defined on a scale of visual entropy in environmentally valid application situations where representation of uncertainty is important for task success?
We demonstrated users could successfully search for glyphs with predefined levels of uncertainty in an urban digital twin visualization of temperature sensors.
\end{itemize}

We believe that visual entropy provides a useful concept with which to reason about glyph shape ordering. We have shown that we can measure it, predict human ranking of glyphs using this measure and apply these  glyphs in a 3D visualization environment. Caveats, as noted, apply to the generality of our results but we believe we have presented a rigorous first step in identifying a new approach for visualizing ordinal data and its application to uncertainty visualization.

\section{Acknowledgement}
The authors wish to thank: the Alan Turing Institute for funding under EPSRC grant EP/N510129/1 and for Professor Holliman’s Turing Fellowship; Professor Jenny Read and Dr Kevin Wilson (Newcastle University) for their helpful insights; Northumbria VRV Studio for the VNG 3D model of Newcastle; the EPSRC UKRIC funded Urban Observatory at Newcastle for sensor data; the Curtin HIVE and members of the CIC at Curtin University, Perth, WA; Dr Ronni Bowman (DSTL) and Matt Butchers (KTN) for organizing inspirational workshops on uncertainty visualization for high level decision makers; all participants who took part in our experiments, and finally, Professor David Firth (University of Warwick) for invaluable advice on statistical methods.

\ifCLASSOPTIONcaptionsoff
  \newpage
\fi



%\bibliographystyle{IEEEtran}
\bibliographystyle{IEEEtran}
\bibliography{references.bib}


\begin{IEEEbiography}
[{\includegraphics[width=1in,height=1.25in,clip,keepaspectratio]{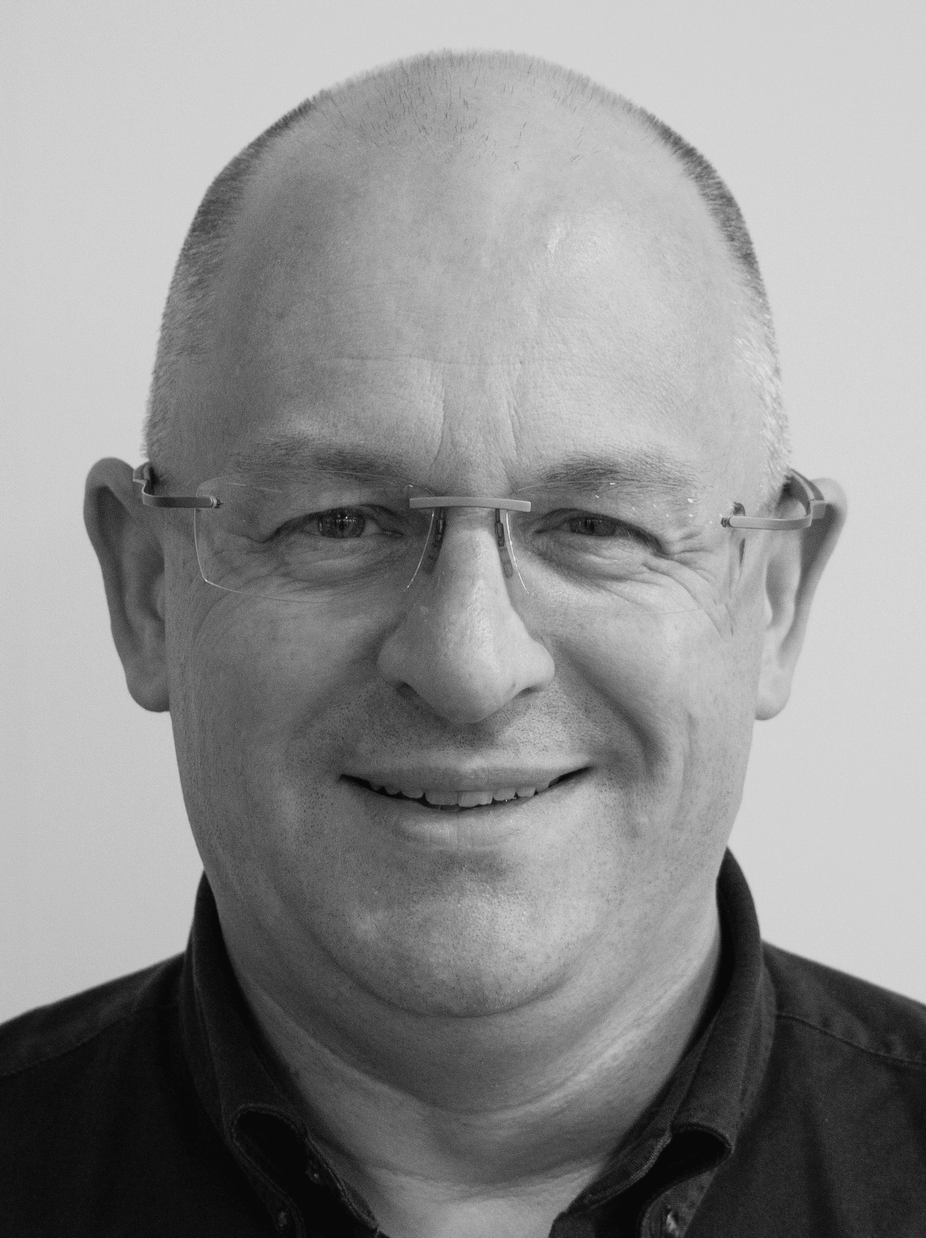}}]
{Nicolas S. Holliman}
Ph.D. (University of Leeds 1990) Computer Science, B.Sc. (University of Durham, 1986) joint hons Computing with Electronics. He has worked in industrial and academic R\&D; was Principal Researcher at Sharp Laboratories of Europe Ltd; Reader in Computer Science at Durham University and then Professor of Interactive Media at the University of York. He is currently Professor of Computer Science at King's College London and is also a Fellow of the Alan Turing Institute in London. He has published in visualization, human vision, computer vision, highly parallel computing and autostereoscopic 3D display systems. He is a member of the IS\&T, the ACM, a fellow of the Royal Statistical Society and a member of the IEEE Computer Society. 
\end{IEEEbiography}

\begin{IEEEbiography}
[{\includegraphics[width=1in,height=1.25in,clip,keepaspectratio]{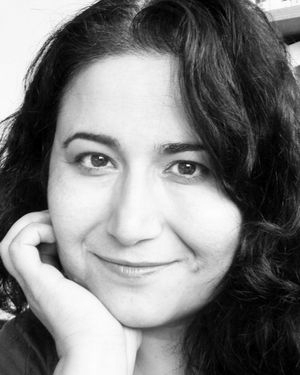}}]
{Arzu Coltekin}
Ph.D. (Helsinki University of Technology, now Aalto, 2006) Photogrammetry and Geographic Information Science. She was a post-doc at the University of Art and Design Helsinki, a group leader at the University of Zurich and is currently a Professor at,  and the director of, the University of Applied Sciences and Arts Northwestern Switzerland's Institute of Interactive Technologies. She chais the international working group on Geovisualization, Augmented and Virtual Reality( ISPRS), co-chairs the Visual Analytics commission (ICA) and is a council member with the ISDE. Her research interests are at the intersection of human-computer interaction, extended reality, visualization, geographic information science and spatial cognition.
\end{IEEEbiography}

\begin{IEEEbiography}
[{\includegraphics[width=1in,height=1.25in,clip,keepaspectratio]{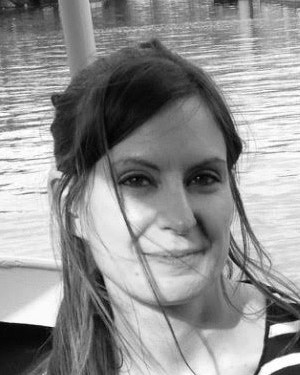}}]
{Sara J. Fernstad}
PhD (Linkoping University, Sweden, 2011) Visualization and Interaction, MSc (Linkoping University, Sweden, 2007) Engineering and Media Technology. She is currently a lecturer in Data Science in the School of Computing at Newcastle University (UK), and previously held a lectureship in Computer Science at University of Northumbria (UK). She carried out post-doctoral research at Unilever R\&D and Cambridge University (UK). Her main research interest is in Information Visualization, with particular interest in visualization of high dimensional data and incomplete data, and the application of visualization to ‘Omics type data.
\end{IEEEbiography}

\begin{IEEEbiography}
[{\includegraphics[width=1in,height=1.25in,clip,keepaspectratio]{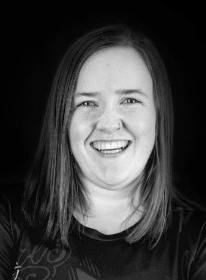}}]
{Lucy McLaughlin}
M.Sc. (Newcastle University, XXXX) , B.Sc. (Newcastle University, XXXX) was an Alan Turing project funded Research Software Engineer and is currently a Ph.D. candidate on the EPSRC funded CDT in Cloud Computing and Big Data at Newcastle University researching novel visualization methods for multivariate and bivariate data. 
\end{IEEEbiography}

\begin{IEEEbiography}
[{\includegraphics[width=1in,height=1.25in,clip,keepaspectratio]{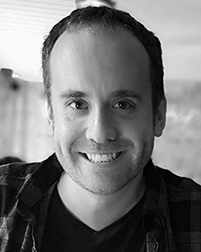}}]
{Michael D. Simpson}
Ph.D. (Newcastle University, 2016) Computing Science, M.Sc. (Newcastle University, 2009) Computer Games Engineering, B.Sc. (Newcastle University, 2008) Computing Science. He worked as a video game developer before returning to complete his Ph.D., applying games industry technology to produce real-time engineering simulations. He is currently part of the Research Software Engineering team at Newcastle University, specializing in the development of interactive 3D applications and visualizations
\end{IEEEbiography}

\begin{IEEEbiography}
[{\includegraphics[width=1in,height=1.25in,clip,keepaspectratio]{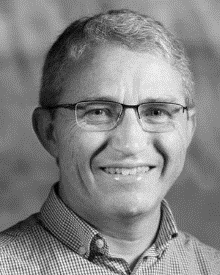}}]
{Andrew J. Woods}
PhD (Curtin), MEng (Curtin), BEng (Curtin). As head of the Curtin University HIVE (Hub for Immersive Visualization and eResearch) he has made contributions to  through his research, development and utilization of 3D technologies. He was recognized as one of Australia’s Most Innovative Engineers for 2017, receiving the honor for his lead role in the development of the deep-water 3D imaging system that was used to survey two of Australia’s most well-known shipwrecks. He is a Fellow of the SPIE and a member of IS\&T, IEAust, IEEE, SID, NSA and ISU.
\end{IEEEbiography}

%\newpage



%\vfill


\begin{figure*}
  \centering
  \includegraphics[width=.95\linewidth]{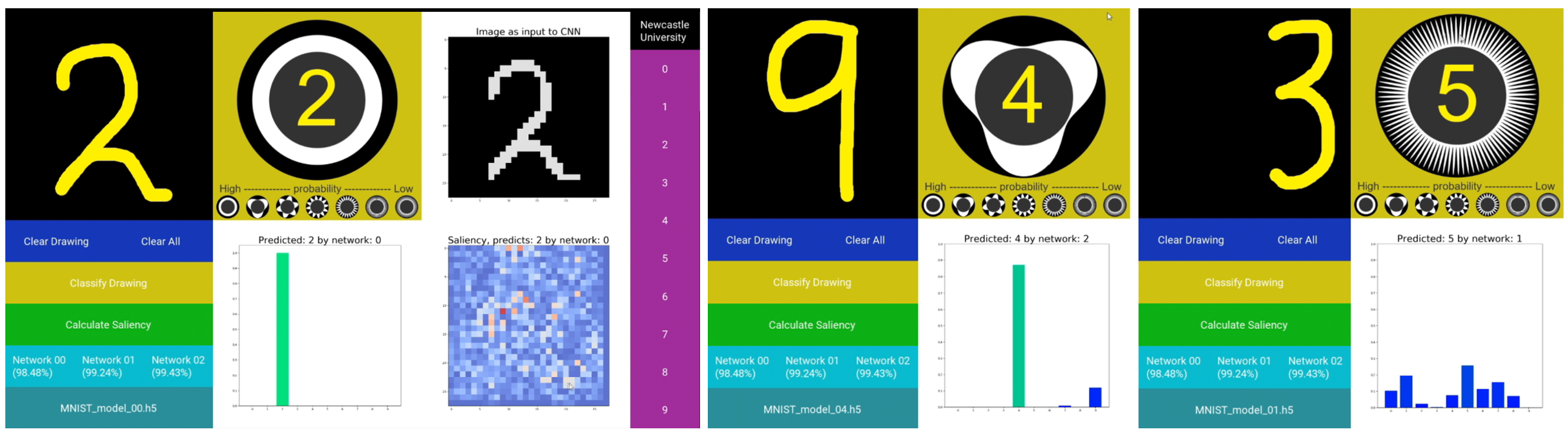}
  \caption{\label{fig.mnist01}
           Appendix: Explainable AI demonstrator: this dashboard allows a user to hand draw a numeric character from 0-9 and then attempts to classify which numeral this is using a convolutional neural network; the probability of each possible outcome is shown in the bar chart and the vizent glyph represents the probability of the most likely classification; the two rightmost panels show failed outcomes and the probability of each using a vizent glyph, the visual entropy shape warning the user the system is less certain than in the first instance. }
\end{figure*}

\appendix
\label{app.glyph}
This technical annex presents additional details on the project beyond the main body of the article.

\subsection{Visual Entropy Glyph Implementation Details for Blender}

The implementation of the glyphs presented in this article used the Blender Cycles path tracer as the scene rendering engine. In normal path tracing the aim is to generate physically representative simulations of light interactions between surfaces, based on their geometry and surface material properties. Blender Cycles is able to do this with high-quality. However, when rendering the areas of our glyphs which carry information as a specified color from a colour map we do not want the physical lighting calculations to be applied as these can distort the color. To implement glyphs so that they work as consistent information representations we had to ensure all colors used in the glyphs were flat shaded with no lighting calculations. One practical way to do this is to make the glyph material highly emissive, shading calculations are then not computed and the glyph is the same color over its entire area independently of the direction of the view camera.

In addition, glyphs are set not to receive or generate shadows, therefore no shadow effects can alter the color appearance, and the glyph shadow cannot alter the appearance of the underlying digital twin (map) layer.
Finally, the glyphs are programmed to rotate to follow the camera position so that the viewer sees a geometrically constant shape for the glyph regardless of their view-point. To do this the entropy glyphs are rotated in two steps. First, the ‘Copy Rotation’ constraint is applied with the Camera as the target, which aligns them with the camera’s rotation. Secondly, the ‘Damped Track’ constraint constraint is applied which rotates the face of the glyph towards the camera. This ensures that the entropy glyph shapes are consistently oriented vertically with the camera to avoid any potential confusion when comparing the shape of glyphs that are spread across the screen.

\subsection{Explainable AI Demonstrator}

One current challenge in AI systems is communicating not just the results produced by the AI system but also the probability of those results. This helps provide the user an indication of the level of trust that they should put in the outcome.

In the demonstrator shown in Fig~\ref{fig.mnist01} we used the Vizent glyph shapes to show probability information for the output classification of a MNIST trained convolutional neural networks (CNN). The dashboard allows a user to hand draw a numeric character from 0-9 and then attempts to classify this using the CNN. We used the demonstrator interactively on a large 85" touch screen with visitors and applicants for courses to demonstrate some of the opportunities and limitations of AI systems.

\subsection{Vizent : an Open Source Implementation using Python}

\begin{figure}[h]
  \centering
  \includegraphics[width=1.0\linewidth]{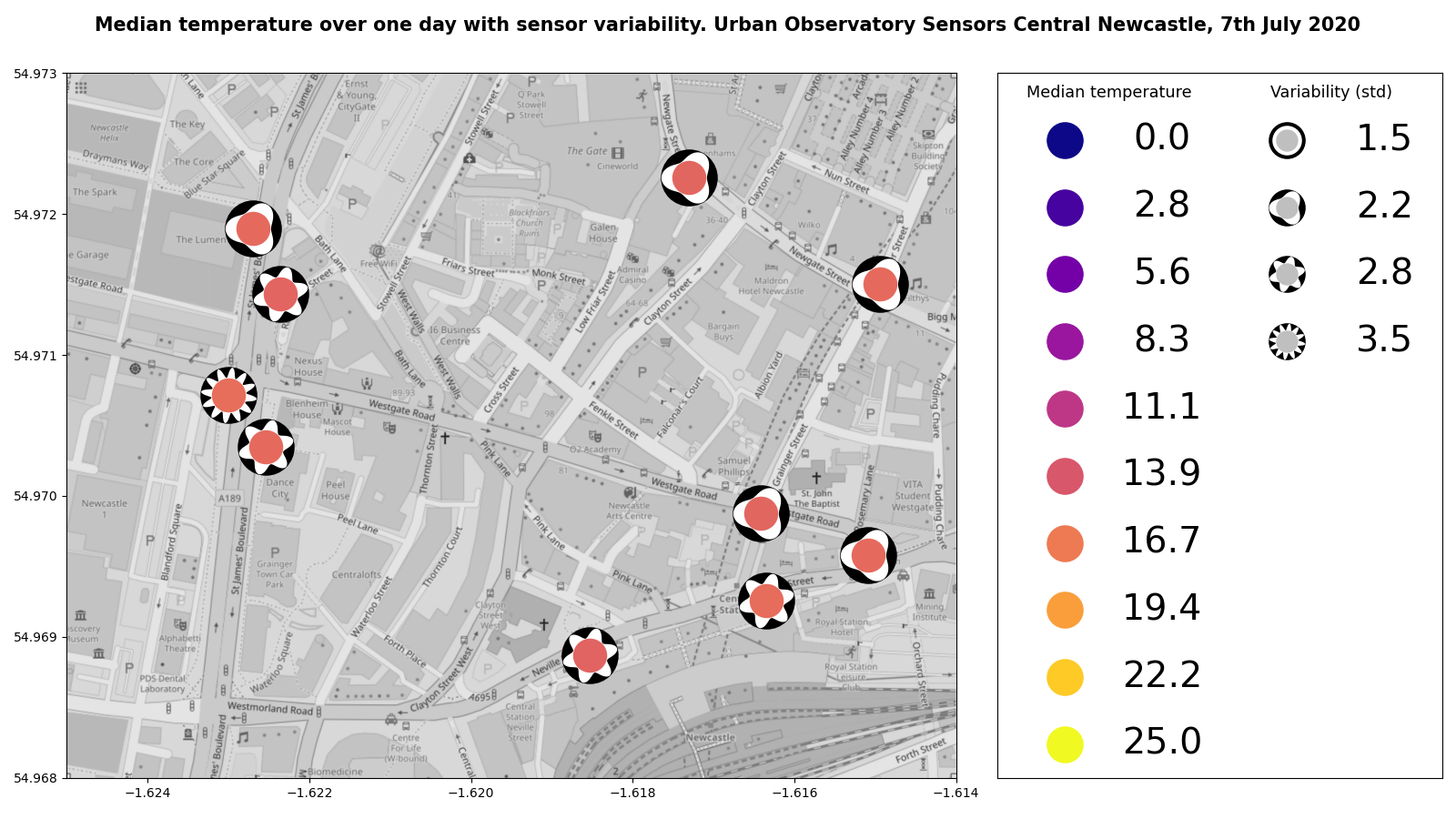}
  \caption{\label{fig.NCLUO01}
           IoT weather data from the Newcastle Urban Observatory visualized using the {\it Vizent} Python library (integrated with Matplotlib and Cartopy) to visualize mean temperature and standard deviation recorded over one day by a set of sensors in central Newcastle in July 2020.}
\end{figure}

We implemented an open source 2D implementation of the visual entropy glyphs, {\it Vizent}, which integrates with the Python Matplotlib library. The {\it Vizent} implementation of the glyphs is available to download from PyPI and Github and a short introductory article is available in the Medium hosted Nightingale magazine by McLaughlin. Fig.~\ref{fig.NCLUO01} demonstrates the use of the {\it Vizent} glyphs in combination with Matplotlib and Cartopy in an application visualizing bivariate IoT data about the local weather in Newcastle-upon-Tyne, England. 

\end{document}